\documentclass[twoside]{article}

\usepackage[sc]{mathpazo}
\usepackage[T1]{fontenc}
\usepackage{microtype}
\usepackage{float}
\usepackage{wasysym}
\usepackage{graphicx}

\usepackage[hmarginratio=1:1,top=32mm,columnsep=20pt]{geometry}
\usepackage[font=it]{caption}
\usepackage{paralist}
\usepackage{multicol}

\usepackage{lettrine}

\usepackage{abstract}

\usepackage{titlesec}
\titleformat{\section}[block]{\large\scshape\centering{\Roman{section}.}}{}{1em}{}

\usepackage{fancyhdr}
\usepackage{hyperref}

\title{\vspace{-15mm}%
	\fontsize{24pt}{10pt}\selectfont
	\textbf{Correlations between Google search data and Mortality Rates}
	}	
\author{%
	\large
	\textsc{James Risk} \\[2mm]
	\normalsize	Michigan State University \\
	\normalsize	\href{mailto:riskjame@stt.msu.edu}{riskjame@stt.msu.edu}
	\vspace{-5mm}
	}
\date{}

\begin{document}

\maketitle
\thispagestyle{fancy}

\begin{abstract}
Inspired by correlations recently discovered between Google search data and financial markets$^1$, we show correlations between Google search data mortality rates. Words with negative connotations may provide for increased mortality rates, while words with positive connotations may provide for decreased mortality rates, and so statistical methods were employed to determine to investigate further.
\end{abstract}

\lettrine[nindent=0em,lines=3]{S} tatistically significant correlations between Google search data and financial markets exist$^1$. The methods used can be used to attempt to find correlations between any time series data and Google$^3$ searches. Since insurance premiums are based on mortality rates, if one can predict mortality rates with a point or interval estimator, an insurance company can re-evaluate a policy every year and update the premium based on what search data may suggest in addition to standard pricing methods.\\

As Google search data and mortality rates are time series data, we can use cross correlations (defined below), to determine if correlations exist and we can show where they exist. Using R, we can use the ccf() (cross-correlation function) command to easily graph and find significant values depending on a significance level (we choose a liberal $\alpha=.1$). In figure 1, we graph a cross correlation function of all races both sexes and the word hate as demonstration. These were the graphs used in the statistical analysis of the project. Figure 2 shows how the significant cross-correlation relates to how the graphs work. \\

Nine words were used for testing. These words are chosen somewhat arbitrarily, but with intentions to hopefully be correlated with mortality. For example, the word "hate" was used and may imply that if more people are hateful (increased search for hate) then mortality rates may increase (due to hate related crimes, for example). \\

In addition, we search nine groups of mortality rates. There is a generic group (all sexes all races) and then groups divided into male and female, and racial groups divided into white and black. The reasoning for this is that it will both let us determine if there is any outlier data, and will also let us see if there are correlations inside of a specific group. It is possible that one race or one sex may react differently to a certain word based on their culture. \\

As discussed above, we present the cross correlation function:
\begin{equation}
\rho_{X, Y} (\Delta t)=\frac{E[X(t)Y(t+\Delta t)]-E[X(t])E[Y(t+\Delta t)]}{\sqrt{E[X^2(t)]-E[X(t)]^2}\sqrt{E[Y^2(t+\Delta t)]-E[Y(t+\Delta t)]^2}}.
\end{equation}
Essentially it checks for relationships between two time series $X(t)$ and $Y(t)$, but at different times, for example if $\Delta t=2$, it will check for $X(t)$'s relationship with $Y(t+2)$'s relationship.  The denominator simply scales the numerator so that $-1 \leq \rho_{X,Y} \leq 1$.

\section{Analysis}
We gather U.S. mortality rates from www.cdc.gov$^2$ and aggregate search data of U.S. citizens from www.google.com$^3$. The mortality data is annual, and Google search data is weekly, so in order to make Google search data annual, we take the yearly average. Then, we want to measure the change in mortality correlated with change in average search volume, because we want to know if next years mortality rates will increase or decrease, and by how much. We then take logs of the differenced data since it is a monotone transformation and reduces variance in the data (the purpose of this paper is to show if correlations exist - further methods can be employed to find how strong they are). \\

\begin{figure}[t]
  \centering
  \includegraphics[width=3in]{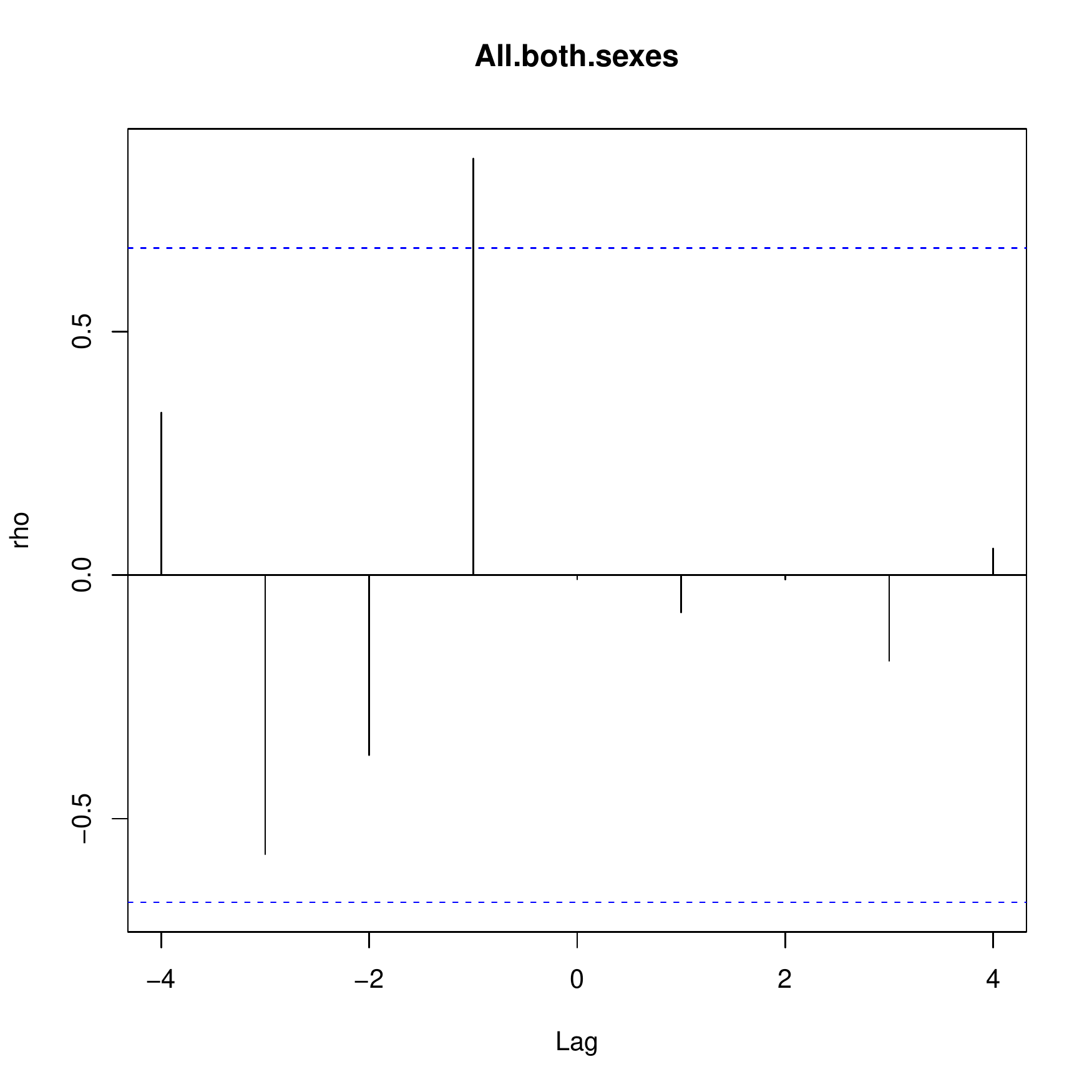}
  \caption{Shows cross correlation function plotted at different lags for all races both sexes and the word hate.  The blue lines are significance intervals at $\alpha=0.1$.  The line outside of the significant band at $\Delta t=-1$ shows that there is a significant correlation at lag $\Delta t=-1$.}
\end{figure}

\begin{figure}[t]
  \centering
  \includegraphics[width=3in]{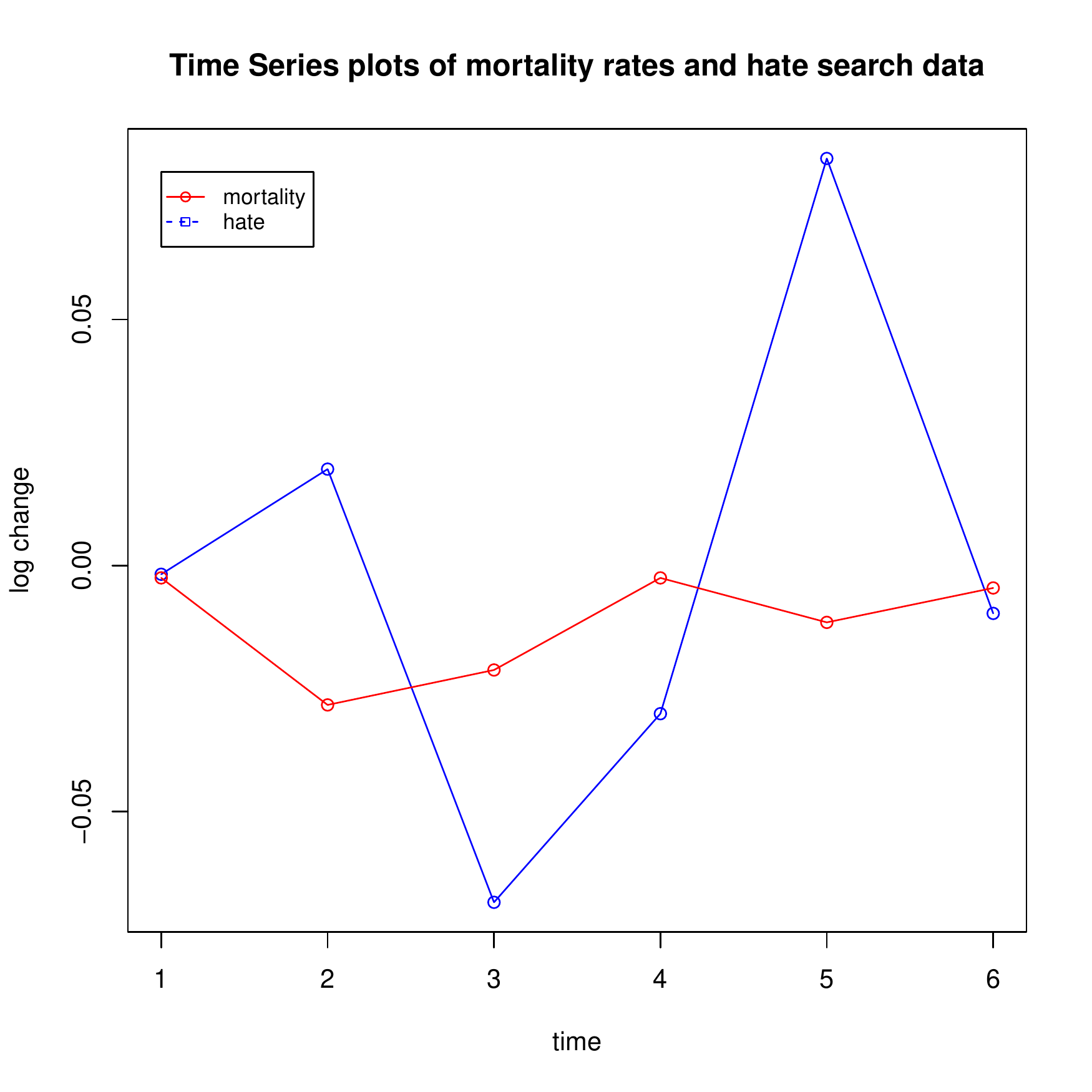}
  \caption{The significant correlation at $\Delta t=-1$ from figure 1 suggests a correlation between how mortality moves and how hate moves one year in the future.  The time series plot checks out with figure 1 and suggests that hate should increase at time 6 to 7, since mortality rate increased from time 5 to 6. }
\end{figure}

Table 1 shows where we found significant cross-correlations and at what lag. An x denotes that there were no significant lags. The significant lags at $\Delta t=-1$ for all people and whites for foreclosure and hate suggests that mortality rates can predict how much people search for those words. The black female data had very strange results, which may suggest that their mortality data in the period that we have is an outlier compared to how their mortality rates normally behave. The only positive lag correlation was with black females and "ship," which, in theory, could suggest that when more people are searching for "ship," more black females are dying, but since the black female data produced strange results, it is most likely that their data is unreliable.  More data should be needed to help with this claim. \\

Unfortunately there are no other significant lags with a positive value, suggesting that none of these words can help predict mortality rates. On the other hand, since for both "foreclosure" and "hate" we found significance at lag $\Delta t=-1$, it may suggest psychological or sociological evidence into how humans react to financial failure and death. Perhaps more people search for "hate" when white people die, but a black person's death has no effect on how often the word "hate" is searched for.

\begin{table}[t]
\centering
\begin{tabular}{|l||c|c|c|c|c|c|c|c|c|}
	\hline
	Word & All & \male & \female & White & Black & White \male & White \female & Black \male & Black \female \\
	\hline
	Marijuana & x & x & x & x & x & x & x & x & x\\
	\hline
	Alcohol & x & x & x & x & x & x & x & x & x\\
	\hline
	Debt & x & x & x & x & x & x & x & x & x\\
	\hline
	Foreclosure & $-1^-$ & $-1^-$ & $-1^-$ & $-1^-$ & $-1^-$ & x & x & x & x\\
	\hline
	Bunnies & x & x & x & x & x & x & x & x & $-3^-$\\
	\hline
	Stress & x & x & x & x & x & x & x & x & x\\
	\hline
	Hate & $-1^+$ & $-1^+$ & $-1^+$ & $-1^+$ & $-1^+$ & $-1^+$ & x & x & $-3^-$\\
	\hline
	Ship & x & x & x & x & x & x & x & $0^-$ & $1^+$\\
	\hline
	Obesity & x & x & x & x & x & x & x & x & x\\
	\hline
	
\end{tabular}
\caption{Table indicating which lags produced statistically significant results.  \male = male, and \female = female.  Where applicable, the superscripts indicate the sign of the correlation.}
\end{table}

\section{Limitations and Suggested Improvements}
The data was very limited which caused many problems. Mortality data was annual, and the latest data point was for 2010. Google search data only went back to 2004, so we had 7 points of data, and differencing these gave us only 6 changes. We are also doing multiple testing at once, so we should account for some sort of test correction to handle the probability of type I error, such as the Bonferroni test correction. There are essentially 81 tests being performed, however, so in order to get some preliminary results we forego these test corrections, as the purpose of this paper is to demonstrate how this method can be used to find correlations and show that there may be significance. In addition, reducing the number of tests is also possible, since we divided up the analysis into several groups races and sexes.  \\

This method could be repeated removing outlier data to possibly improve the result. For example, the black women data acted strangely, which may suggest the data we had for this time period did not follow normal mortality trends. If we remove black women from the overall mortality data, we may get more accurate conclusions. \\

In addition, in the future our analytical methods are more practical. For example, 10 years from now we will have more than double the data points we had previously, causing for a huge improvement in accuracy and potential results. In addition, if more detailed mortality data exists, we can test for correlations between certain words and age groups. As an example, we could find positive correlation between people searching for "alcohol" and deaths for people in the age range of 15-25. \\

Although it will probably not be possible, if mortality data was available weekly, it would be tremendous for data analysis, since we will have 52 times as many data points.

\section{Conclusion}
We did find significant evidence at negative lags even though our data provided no significant conclusions for positive lag cross-correlations, This opens two new doors. The first being that maybe we haven't found the right word, and given the correct word we can predict mortality rates. The second being that we did find significant evidence for negative lags, providing psychological or sociological insight as to how mortality rates affect our searching on Google. Further work can definitely be done to find correlations with mortality, but even aside from that it provides more evidence as to how our lives affects Google search data and how Google search data affects our lives.

\section{References}
\begin{enumerate}[1.]
	\item Preis, T., D. Reith, and H. E. Stanley. "Complex Dynamics of Our Economic Life on Different Scales: Insights from Search Engine Query Data." \textit{Philosophical Transactions of the Royal Society A: Mathematical, Physical and Engineering Sciences} 368.1933 (2010): 5707-719. Print.
	\item \textit{Centers for Disease Control and Prevention.} Centers for Disease Control and Prevention. Web. 27 Apr. 2012. <http://www.cdc.gov/>.
	\item "Google Insights for Search." \textit{Google.} Web. 27 Apr. 2012.\\ <http://www.google.com/insights/search/>.
\end{enumerate}

\end{document}